\documentclass[11pt]{article}
\pdfoutput=1
\usepackage{graphicx,amsmath,amssymb,amsfonts}
\usepackage{enumerate,setspace}
\usepackage{psfrag}
\usepackage{color}   
\usepackage{rotating}
\usepackage{pbox}
\usepackage{framed}
\usepackage{geometry}
\usepackage{floatrow}
\usepackage{bm}        
\usepackage{color}
\usepackage{multirow}
\usepackage{color}
\usepackage{relsize}
\usepackage{hyperref}
\usepackage{bigints}
\usepackage{setspace}
\usepackage{times}
\definecolor{darkblue}{rgb}{0,0,.5}

  \usepackage{natbib}
  \usepackage{pdfpages}
\usepackage{hyperref}
  
\newcommand{\EQ}{\begin{equation}}
\newcommand{\EE}{\end{equation}}
\newcommand{\EQA}{\begin{eqnarray}}
\newcommand{\EEA}{\end{eqnarray}}
\usepackage[labelfont={bf,color=black},font={small}]{caption}

\usepackage{makeidx}

\renewcommand{\small}{\footnotesize}

\newcommand{\ol}{\overline}
\usepackage{diagbox}

\newcommand{\comment}{}

\renewcommand{\ol}[1]{{\overline{ #1}}}
\begin{document}

\begin{titlepage}
\title{Fierce selection and interference in B-cell repertoire response to  chronic HIV-1}
\author{Armita Nourmohammad$^{1,2*}$, Jakub Otwinowski$^{1}$, Marta \L uksza$^{3}$, \\  Thierry Mora$^{4\dagger}$, Aleksandra M Walczak$^{5\dagger}$}
\date{\small $^1$ Max Planck Institute for Dynamics and Self-organization, Am Fa\ss berg 17, 37077 G\"ottingen, Germany\\ 
$^2$Department of Physics, University of Washington, 3910 15th Avenue Northeast, Seattle, WA 98195, USA\\ 
$^3$Tisch Cancer Institute, Icahn School of Medicine at Mount Sinai, 1470 Madison Ave, New York, NY 10029, USA\\
$^4$Laboratoire de Physique Statistique, CNRS, Sorbonne University, Paris-Diderot University, \'Ecole Normale Sup\'erieure (PSL), 24, rue Lhomond, 75005 Paris, France\\
$^5$Laboratoire de Physique Th\'eorique, CNRS, Sorbonne University, \'Ecole Normale Sup\'erieure (PSL), 24, rue Lhomond, 75005 Paris, France\\
}
\maketitle

\noindent{$^*$ correspondence should be addressed to Armita Nourmohammad: armita@ds.mpg.de}\\
\noindent{$\dagger$ equal contribution}\\

\noindent
{{\bf Abstract} \\
During chronic infection, HIV-1 engages in a rapid coevolutionary arms race with the host's adaptive immune system. While it is clear that HIV exerts strong selection on the adaptive immune system, the characteristics of the somatic evolution that shape the immune response are still unknown. Traditional population genetics methods fail to distinguish chronic immune response from healthy repertoire evolution. Here, we infer the evolutionary modes  of B-cell repertoires and identify complex dynamics with a constant production of better B-cell receptor mutants that compete, maintaining large clonal diversity and potentially slowing down adaptation. A substantial fraction of mutations that rise to high frequencies in pathogen engaging CDRs of B-cell receptors (BCRs) are beneficial, in contrast to many such changes in structurally relevant frameworks that are deleterious and circulate by hitchhiking.  We identify a pattern where BCRs in patients who experience larger viral expansions undergo stronger selection with a rapid turnover of beneficial mutations due to clonal interference in their CDR3 regions. Using population genetics modeling, we show that the extinction of these beneficial mutations can be attributed to the rise of competing beneficial alleles and clonal interference. The picture is of a dynamic repertoire, where better clones may be outcompeted by new mutants before they fix.}
   
\end{titlepage}

\section*{Introduction}

HIV-1 evolves and proliferates quickly within the human body~\citep{Richman:2003dc,Moore:2009hv,Liao:2013gs}, rapidly mutating and often recombining its genetic material among different viral genomes. These factors make it very hard for the host immune system to maintain a sustained control of an infection, leading to a long-term chronic condition.  {While it is  clear that the virus exerts strong selective pressure on the host immune system, the quantitative nature of the evolutionary dynamics of the  adaptive immune  system during chronic infections remains unknown. }

\begin{figure}[t!]
\centering \includegraphics[width=\textwidth]{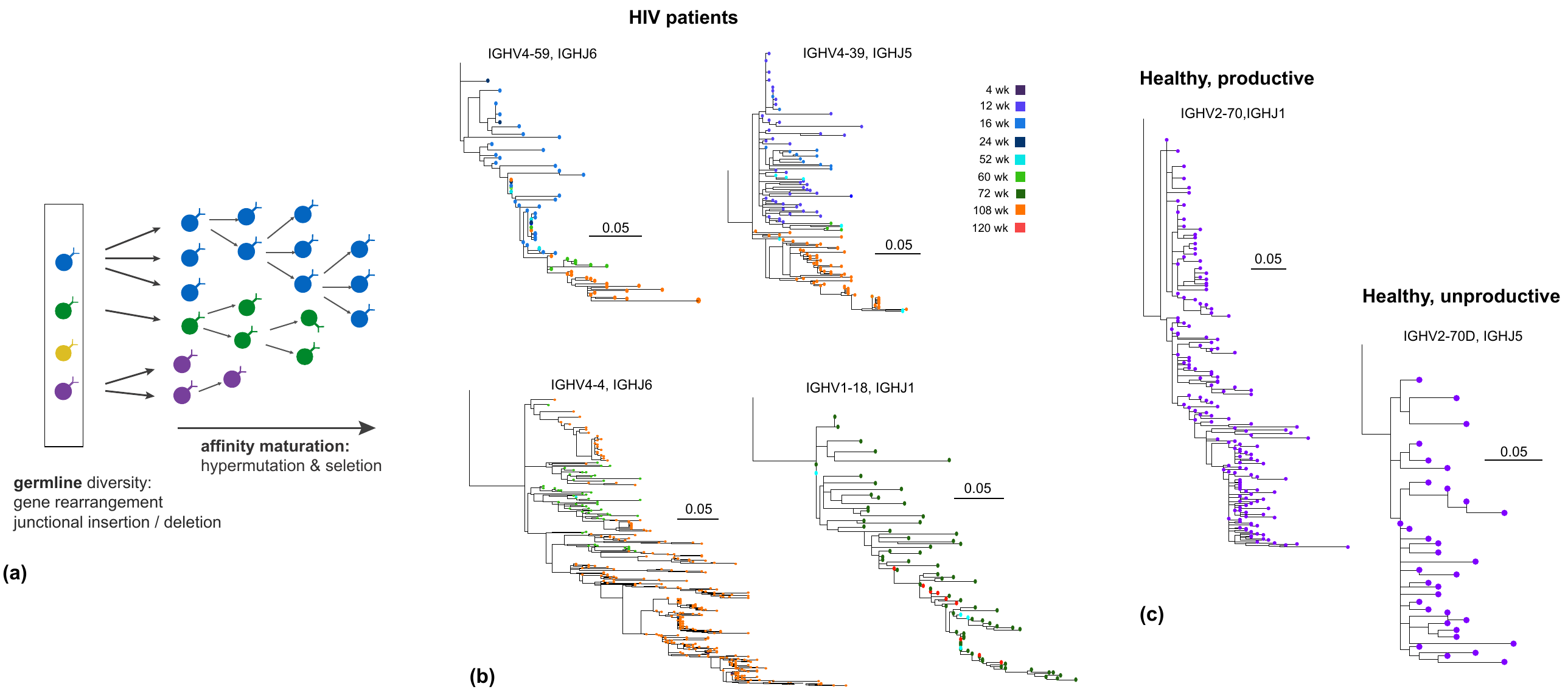}
\caption{
{{ \bf Affinity maturation forms B-cell lineages.} {\bf (a)}  Schematic of B-cell affinity maturation and lineage formation.  The naive immune repertoire consists of a diverse set of B-cell receptors, generated by gene rearrangement (VDJ recombination) and  junctional  sequence insertion and deletion (distinct colored cells in the box). Affinity maturation with somatic hypermutations and selection for strong binding of BCRs to antigens forms lineages of BCRs stemmed from a germline progenitor, shown by three growing lineages in this figure. {\bf (b)} Examples of B-cell lineages reconstructed from the heavy chain sequences of  BCR repertoires in HIV patients (see Methods and SI). The distance between the nodes along the horizontal axis indicates their sequence hamming distance. The nodes are colored according to the time they were sampled from a patient over the period of $\sim 2.5$ yrs. {\bf (c)}  Examples of a productive (left) and unproductive (right) B-cell lineage reconstructed  from the heavy chain repertoire of a healthy individual sampled  at a single time point (Methods, SI).
}
\label{fig:Fig1}}
\end{figure}

The immune system has a diverse set of B and T-cells with specialized surface receptors that recognize foreign antigens, such as viral epitopes, to protect the organism. We focus on the chronic  phase of HIV infection, where the immune response is dominated by antibody-mediated mechanisms, following the strong response of cytotoxic T-lymphocytes (i.e., CD8$+$ killers T-cells), around 50 days after infection~\citep{McMichael:2010cg}.  During the chronic phase, the symptoms are minor and the viral load is  relatively stable, but its  genetic composition undergoes rapid turnover.   After an infection, B-cells undergo a rapid somatic  hypermutation in lymph node germinal centers, with a rate that is approximately $4-5$ orders of magnitude larger than an average germline mutation rate in humans~\citep{Campbell:2013cp}. Mutated B-cells compete for survival and proliferation signals from helper T-cells, based on the B-cell receptor's binding to antigens. This process of \emph{affinity maturation} is Darwinian evolution within the host and can increase binding affinities of B-cell receptors (BCRs) up to 10-100 fold~\citep{Victora:2012gx}.  It generates memory and plasma  B-cells with distinct receptors,  forming lineages that reflect their co-evolution with viruses~\citep{NourMohammad:2016hg},  (see schematic in Fig.~\ref{fig:Fig1}a). A B-cell repertoire consists of many  such lineages forming a forest of co-existing genealogies. {The outcome of an affinity maturation process shifts the overall repertoire response against the pathogen~\citep{Berek:1987tf}.}

Immune repertoire high-throughput sequencing has been instrumental in quantifying the diversity of B-cell repertoires~\citep{Weinstein:2009he,Elhanati:2015cp}. Statistical methods have been developed to characterize the  processes involved in the generation of diversity in repertoires and to infer the underlying heterogenous hypermutation preferences in B-cell receptors~\citep{Yaari:2013ea,Elhanati:2015cp,McCoy:2015ga}.  Deviation of the observed mutations in BCRs from the expected hypermutation patterns are used to infer selection effects of mutations from repertoire snapshots in order to identify functional changes that contribute to the response against pathogens~\citep{Yaari:2013ea,Uduman:2014bg}. Recently, longitudinal data, with repertoires sampled over multiple time points from the same individuals, have brought insight into the dynamics of affinity maturation in response to antigens~\citep{Vollmers:2013bm,Laserson:2014jo,Hoehn:2015kl,Horns:2017kl}.  The dynamics of affinity maturation and selection in response to HIV have also been characterized for chosen monoclonal broadly neutralizing antibody lineages~\citep{Liao:2013gs,Vieira:2017ft}.  Yet, the effect of a chronic infection on the dynamics of the whole BCR repertoire remains unknown.
  
{Here, we analyze the history and structure of BCR lineages in the full repertoire of HIV-1 infected patients. We uncover distinct modes of immune response, including selection and  competitive clonal interference  among BCRs, a fraction of which may be HIV-specific. We identify a pattern, where  BCR repertoires in patients who experience a larger viral expansions undergo stronger selection and clonal interference in their pathogen-engaging CDR3 regions. We show that  clonal interference in CDR3  regions reflects a macro-evolutionary drive of the repertoire, either caused by the virus or the overall reorganization of the BCRs, even those that do not directly target HIV-1. Our results are based on advanced statistical measures informed by population genetics theory that capture the differences between baseline affinity maturation and long-term selection in response to HIV-1 infection.
}

\section*{Results} 
We compare the structure and dynamics of BCR repertoires sampled over 2.5 years in HIV patients (data from ref.~\citep{Hoehn:2015kl}, collected through the SPARTAC study~\citep{SPARTACTrialInvestigators:2013dw}). Among these individuals are  2 untreated patients  and 4 patients who had interrupted  ART after a year of treatment. We have also analyzed the   BCR repertoire structure in 3 healthy individuals  (data from ref.~\citep{DeWitt:2016cp}). The sequencing depth of the two datasets differ, with on average  $\sim 172,000$  unique BCR sequences per HIV patients,  and $\sim 880,000$ unique BCRs in healthy individuals and an average of about 3,500 lineages with size $>20$ per HIV patient and 17,700 per healthy individuals; see Methods, SI, Fig.~S1 and Table~S1 for details on BCR data and processing.
Additionally, \comment{due to the differences in the sequencing protocols~\citep{Hoehn:2015kl,DeWitt:2016cp}, the read length of  the receptors} in healthy individuals ($\sim 130$ bp)  is much smaller than in HIV patients ($\sim 300$ bp with $\sim 35$ bp gap), making a direct comparison between  the two datasets difficult. We have performed  our statistical analysis  both on the complete BCR repertoire data in healthy individuals and  on the sub-sampled data with a depth comparable to the BCR repertoires in HIV patients; see SI. However, the healthy repertoires serve as a  guideline in our analysis, rather than a null model for selection in chronically challenged BCR repertoires, due to the differences in the structure of the datasets and the underlying sequencing protocols. Our primary conclusions rely on the  analysis  of selection in BCR repertories of HIV patients and relating the differences  among patients to  the state of their viral load over time.

\subsection*{Statistics of BCR lineage genealogies indicate positive selection}
\begin{figure}[t!]
\centering\includegraphics[width=.6\textwidth]{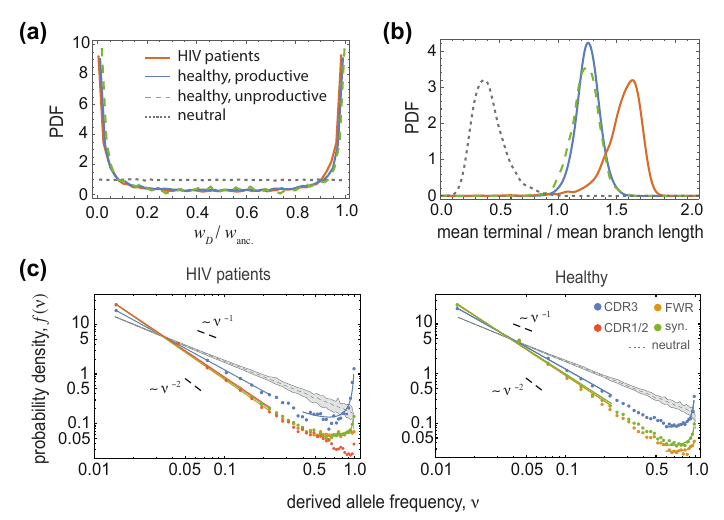}
\caption{\label{fig:Fig2}
{{\bf Statistics of BCR lineage genealogies indicate positive selection.} {\bf(a)} The U-shaped distribution of sub-lineage weight ratios at the root of lineage trees (SI)  $ w_{D}/w_{\rm anc.}$  and {\bf (b)} the distribution of elongated mean  terminal branch length (in units of divergence time) relative to  the mean length of all branches in  BCR lineages  indicate positive selection in HIV patients and in healthy individuals (colors), in contrast to the neutral expectation (dotted lines); see Fig.~S2 for comparison of tree statistics  under different evolutionary scenarios.   {\bf (c)} The Site Frequency Spectrum (SFS) $f(\nu)$ is shown for mutations  in different regions of BCRs (distinct colors) in HIV patients (left) and in healthy individuals (right); see Fig.~S3 for SFS of  unproductive BCR lineages. {The frequencies are estimated within each lineage and the distributions are aggregates across lineages of size $>100$, amounting to a total of 1524 lineages in HIV patients and 2795 lineages in healthy individuals; see Table~S1 for details.   The grey area shows the span of SFS across 100 realizations of simulated neutral trees (Kingman's coalescent) with  sizes equal to the  BCR lineages in HIV patients (left) and in healthy individuals (right). 
 The significant upturn of the SFS for non-synonymous mutations in the CDR3 region  is indicative of rapid evolution under positive selection. The upturn for synonymous mutations indicates hitchhiking of neutral mutations along with the  positively selected alleles (SI).}
}}
\end{figure}

{We reconstruct genealogical trees for  B-cell receptor lineages inferred from BCR repertoires in each individual (Methods, SI).  B-cell lineages of HIV patients, a few examples of which are shown in Fig.~\ref{fig:Fig1}b, can persist from months to years after the initial infection, which is much longer than the lifetime of a germinal centre (weeks), indicating the recruitment of memory cells for further cycles of affinity maturation in response to the evolving virus.} Reconstructed lineage trees show a skewed and asymmetric structure, consistent with rapid evolution under positive selection (see Fig.~S2)~\citep{Neher:2013il}.  To quantify these asymmetries, we estimated two indices of tree imbalance and terminal branch length anomaly. In both HIV patients and healthy individuals, we observe a significant branching imbalance  at the root of the BCR lineage trees, indicated by the U-shaped distribution of the sub-lineage weight ratios (see SI), in contrast to the flat prediction of neutral evolution, calculated from Kingman's coalescent (Fig.~\ref{fig:Fig2}a). Moreover, we observe elongated terminal branches {(i.e., larger coalescence time)} in BCR trees compared to their internal branches, with the strongest effect seen in trees from HIV patients, again in violation of neutrality (Fig.~\ref{fig:Fig2}b, Fig.~S2); {see SI for inference of coalescence time}.  These asymmetric features of BCR trees are clear signs of intra-lineage positive selection. {They break the assumptions of neutral models that are based on non-biased growth of all terminal branches, which results in all branches and sub-lineages growing at equal rates. However, the considered statistics } only reflect the history of lineage replication and give limited insight into the mechanisms and dynamics of selection. For instance, tree asymmetry is also observed in unproductive BCR lineages, which lack any immunological function but are carried along with the productive version of the recombined gene expressed on the other chromosome (Fig.~\ref{fig:Fig2}a,b).\\
 
\subsection*{Site frequency spectra indicate  rapid adaptation in CDR3 regions}
To characterize the selection effect of mutations in more detail, we evaluate the spectrum of mutation frequencies in a lineage,  known as the  site frequency spectrum (SFS). {The SFS is the probability density $f (\nu)$ of observing a derived mutation (allele) with a given frequency $\nu$ in a lineage. A mutation that occurs along the phylogeny of a lineage forms a clade and is present in all the descendent  nodes (leaves) of its clade (see  Fig.~S2). Therefore, SFS carries information about the shape of the phylogeny, including both the topology and the branch lengths. In  neutrality,  mutations rarely reach high frequency, and hence, the SFS decays monotonically with allele frequency as, $f(\nu)\sim \nu^{-1}$~\citep{Kingman:1982jap}. In phylogenies with skewed branching, many mutations reside on the larger sub-clade following a branching event, and hence, are present in the majority of the descendent leaves on the tree. The SFS of such lineages is often non-monotonic with an upturn in the  high frequency part of the spectrum~\citep{Neher:2013il}}. We evaluate the SFS separately for synonymous and non-synonymous mutations in different regions of BCRs (Fig.~\ref{fig:Fig2}c, Fig.~S3 and SI). In HIV patients, we see a signifiant upturn of SFS polarized on non-synonymous mutations in pathogen-engaging CDR3 regions, consistent with rapid adaptive evolution~\citep{Neher:2013il}, and in contrast to monotonically decaying SFS in neutrality (Fig.~\ref{fig:Fig2}c, SI). {In addition, we observe significant over-representation  of high-frequency synonymous mutations in productive lineages of  HIV patients and healthy individuals, which indicates hitchhiking of neutral mutations with positively selected alleles. We evaluate the significance of the signal by comparing to a bootstrapped distribution of an ensemble of neutrally generated trees with otherwise similar statistics to experimentally observed  BCR lineages (Fig.~\ref{fig:Fig2}c, Fig.~S3 and SI).}   The signal of positive selection is strongest in HIV patients with an order of magnitude increase in the high end of the spectrum, suggesting that the BCR population rapidly adapts in  HIV patients. {In addition, this signal is not an artifact of heterogenous hypermutation patterns in BCRs, as shown by simulations in SI and  in Fig.~S4. }

{A similar signal of adaptation based on the upturn in the site frequency spectrum has been observed among  BCR lineages in response to influenza vaccine in healthy individuals~\citep{Horns:2017kl}. Although the upturn of SFS is often used as a standard signal for selection in population genetics, it has low power in distinguishing between hitchhiking under selection or out-of-equilibrium effects due to population structure in neutrality~\citep{Jensen:2005} (SI). In particular, the signal may be confounded in expanding populations of B-cells during transient response  to acute infections or vaccination.}\\

\begin{figure}[t!]
\begin{center}\includegraphics[width=0.8\textwidth]{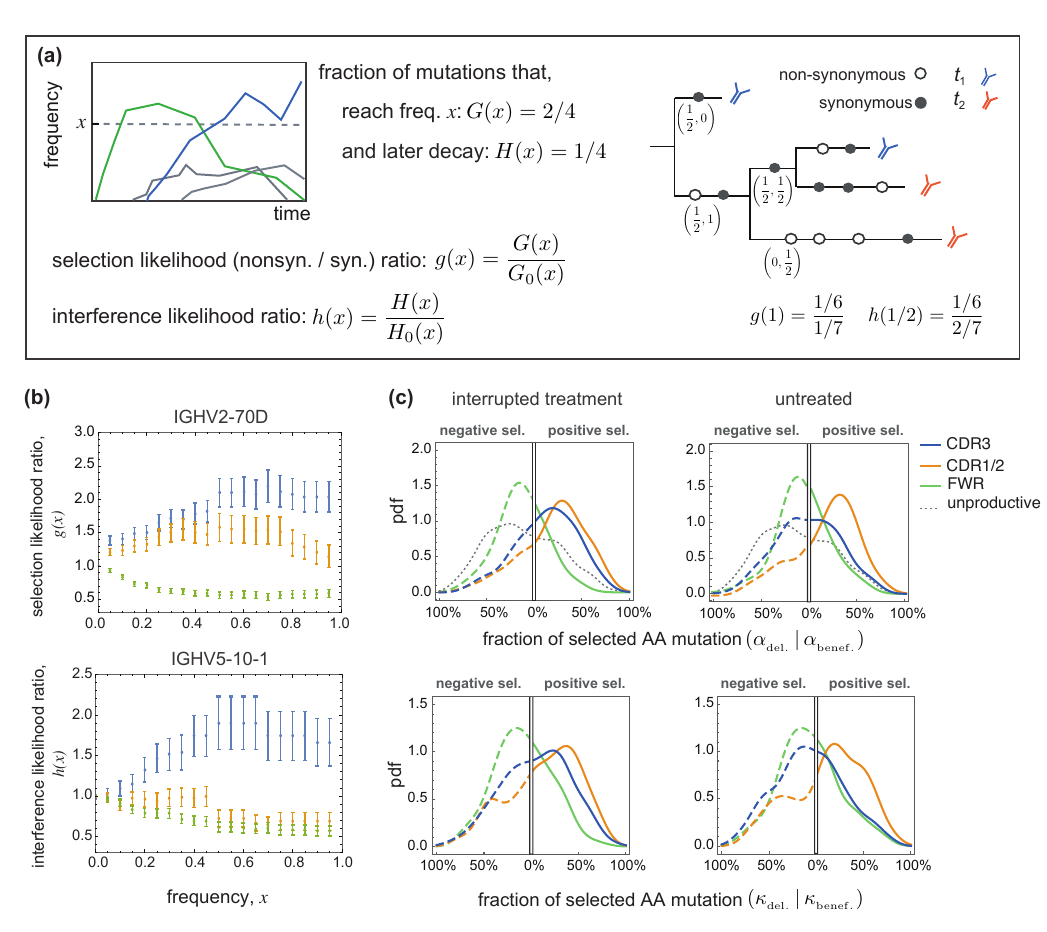}\end{center}
\caption{\label{fig:Fig3}
{{ \bf Inference of selection and clonal interference in BCR lineages.} {\bf (a)} 
Schematic shows  time-dependent frequencies of { four distinct mutations} that rise within a population (left). We denote the fraction of mutations that reach frequency $x$ {(at any time point) } within a population (or lineage) by $G(x)$  (blue and green) and the subset that later goes extinct due to clonal interference by $H(x)$ (green). {Classifying the mutations into  non-synonymous and synonymous groups, we quantify the strength of selection using the likelihood ratios between the non-synonymous and synonymous mutations, $g(x) ={G(x)}/{G_0(x)}\text{, and } h(x) = {H(x)}/{H_0(x)}$.}  A schematic B-cell genealogy (right) is shown for a lineage sampled at two time points (colors).  Non-synonymous and synonymous mutations are shown by empty and filled circles and their {time-dependent frequencies  }$(x_{t_1},x_{t_2})$, as observed in the sampled tree leaves,  are indicated below a number of branches. {The corresponding likelihood ratios are given below.} {\bf (b)} Selection likelihood ratio $g(x)$ in the V-gene class IGHV2-70D (top;  {pooled from 35 lineages}) and the interference  likelihood ratio $h(x)$ for the V-gene class IGHV5-10-1(bottom;  {pooled from 18 lineages}) in patient 5 are plotted against frequency $x$ for mutations in different BCR regions (colors). The likelihood ratios indicate positive selection and strong clonal interference in the CDR3 region, negative selection on the FWR region and positive selection on mutations that rise to intermediate frequencies  in the joint CDR1 / CDR2 regions. We do not observe interference in the FWR and the joint CDR1 / CDR2 region. The error bars are estimated assuming a binomial sampling of the mutations  (see SI). {\bf (c)} {Each panel shows the probability density across distinct VJ-gene classes  in HIV patients with interrupted treatment (left) and without treatment (right), of the fractions of beneficial and deleterious mutations  $\alpha_{{}_{\rm benef.}}$ and $\alpha_{{}_{\rm del.}}$ on the right x-axis and the left inverted x-axis, respectively that reach frequency $x= 80\%$ within a lineage (top), and similarly, the fractions of beneficial and deleterious mutations   ($\kappa_{{}_{\rm benef.}}$ and $\kappa_{{}_{\rm del.}}$) that reach frequency $x= 60\%$ within a lineage and later go extinct  (bottom). The fraction of selected mutations $\alpha,\, \kappa$ are estimated based on the deviation of the  likelihood ratios, $g(x)$ and $h(x)$, from 1 in  VJ-gene classes within each  patient separately (Methods, SI). These aggregate statistics are then pooled together to form the histograms, without averaging over VJ-genes across patients. The dotted grey line indicates the null distribution from unproductive lineages of healthy individuals (Fig.~S8). The probability densities are evaluated from 13,601 lineages and  aggregated over 661 VJ-gene classes pooled from the 4 patients with interrupted treatment (left), from 7,043 lineages with 373 VJ-gene classes pooled from the 2 untreated  patients (right) and from 2,903 unproductive lineages with 417 VJ-gene classes pooled from 3 healthy individuals (dotted grey).} The color code for distinct BCR regions  in all panels is consistent with the legend; see SI for statistical details, Table~S1 for details on lineages and Fig.~\ref{fig:Fig4}, Figs.~S5,~S7,~S8,~S10 for further analysis of  likelihood ratios and selection statistics.
 }}
\end{figure}

\subsection*{Inferring intra-lineage selection and interference from longitudinal data}

To understand the dynamics and fate of  adaptive mutations during chronic infection, we use the longitudinal nature of the data to analyze the temporal structure of the lineages. We estimate the likelihood that a new mutation appearing  in a certain region of a BCR reaches frequency $x$ at some later time  within the lineage (Fig.~\ref{fig:Fig3}a), and evaluate a measure of selection  $g(x)$ as the ratio of this likelihood between non-synonymous and synonymous mutations~\citep{Strelkowa:2012jo} (Methods, SI).  {The frequency of a mutation $x_t$ is estimated as the relative size of its descendent clade at time $t$ (number of leaves in its sub-clade ) to the total number of  leaves in the lineage at that time  (Fig.~\ref{fig:Fig3}a). } At frequency $x=1$ (i.e., substitution), the likelihood ratio $g(x)$ is  equivalent to the McDonald-Kreitman test for selection~\citep{Kreitman:1991vh}.  Generalizing it to $x<1$ makes it a more flexible measure applicable to the majority of mutations  that only reach intermediate frequencies. {Similar to McDonald-Kreitman test, the likelihood ratio $g(x)$ is  relatively  robust to effects due to demography in comparison to the SFS, as both synonymous and non-synonymous mutations experience similar demographic biases.}

A major reason why many beneficial mutations never fix  in a lineage is clonal interference, whereby  BCR mutants within and across lineages compete with each other~\citep{NourMohammad:2016hg}. Clonal interference in population genetics refers to a specific regime of evolution by natural selection, where multiple beneficial mutations simultaneously and independently arise on different genetic backgrounds and form competing clones. Here, we use the population genetics definition of a  ``clone", which refers to the descendants (i.e., sub-clade) of  a given mutation in a lineage phylogeny, and although related, it should not be confused by the immunological analogue in ``clonal selection theory"~\citep{Burnet:1976uu}. In the absence of  clonal interference, beneficial mutations can readily fix after they rise to intermediate frequencies, beyond which  stochastic effects cannot  impact their fate~\citep{Desai:2007wv} (Methods, SI). \comment{Clonal competition among beneficial mutations is common in large adaptive asexual populations and reduces the rate of evolution  by slowing down the  successive fixation of beneficial mutations~\citep{Schiffels:2011fua}. In this evolutionary regime, the  dynamics of  beneficial mutations becomes more neutral~\citep{Schiffels:2011fua}, resulting in a reduced  efficacy of selection that hinders the emergence of very fit strains (e.g. a high affinity BCR). Moreover, the non-linearity due to  competition among clones reduces the predictability of  the fate of  beneficial mutations during evolution~\citep{Lassig:2017hr}. }

To quantify the prevalence of clonal interference, we evaluate the non-synonymous-to-synonymous ratio $h(x)$ as the likelihood for a mutation to reach frequency $x$ and later to go extinct~\citep{Strelkowa:2012jo} (Fig.~\ref{fig:Fig3}a, Methods and SI). In short, the selection likelihood $g(x)$ identifies ``surges'' and interference likelihood $h(x)$ ``bumps" in frequency trajectories of clones. These likelihood ratios have intuitive interpretations: $g(x)>1$ {shows over-representation of non-synonymous to synonymous mutations at frequency $x$  and} indicates evolution under positive selection, with a fraction of at least $\alpha_{{}_{\rm benef.}}(x)=1-1/g(x)$ strongly beneficial amino acid mutations in a given region~\citep{Smith:2002cm}. On the other hand, the likelihood ratio $g(x)$ smaller than 1 is indicative of negative selection{, where non-synonymous mutations are suppressed},  with a fraction of  at least $\alpha_{{}_{\rm del.}}(x)=1-g(x)$  strongly deleterious mutations  (see SI for a derivation of these bounds). Likewise, $\kappa_{{}_{\rm benef.}}(x)=1-1/h(x)$ or $\kappa_{{}_{\rm del.}}(x)=1-h(x)$ define a lower bound on the fraction of either beneficial or deleterious mutations that go extinct.\\

\subsection*{Region specific patterns of intra-lineage selection and interference}

To demonstrate the structure of the signal, Fig.~\ref{fig:Fig3}b shows the selection likelihood ratio $g(x)$ in an HIV patient  (patient 5) for  lineages belonging to a typical V-gene class IGHV2-70D (Methods); see Fig.~S5 for repertoire averaged statistics in all individuals. In this gene family, we detect positive selection ($g>1$) in the CDR3 region. {We observe about $10\%$ of the 854 non-synonymous mutations in this gene-family  reach frequency $x=0.6$, in comparison to only $5\%$ of the 884 synonymous mutations;  mutations are pooled across 35 lineages with an average CDR3 length of 45 bp. Therefore, the selection likelihood ratio $g(x)$ has around a two fold larger fraction of non-synonymous compared to synonymous mutations in the CDR3 region, which  indicates that at least  $\alpha_{{}_{\rm benef.}}=46\%$ of mutations that reach frequency $x=0.6$ are strongly beneficial. On the other hand, the likelihood ratio in  FWR  signals strong negative selection ($g<1$), where non-synonymous mutations  reaching frequencies $x=0.6$ are two times less frequent than the synonymous mutations, which indicates at least $\alpha_{{}_{\rm del.}}=37\%$ of these mutations are strongly deleterious.} {In FWR we identify $3987$ non-synonymous mutations, $2\%$ of which reach frequency $x=0.6$, in comparison to $4\% $ of the 3114 synonymous mutations; the average FWR length among the 35 pooled lineages   is 213 bp.}  Similarly, the interference likelihood ratio $h(x)$ for a V-gene class IGHV5-10-1 in patient 5 indicates that at least $\kappa_{{}_{\rm benef.}}=37\%$ of CDR3 mutations in this gene family that go extinct due to clonal interference are strongly beneficial (Fig.~\ref{fig:Fig3}b).
{ This likelihood ratio is estimated based on the observed  $16 \%$ of the  231 non-synonymous mutations that reach frequency $x=0.6$ and later go extinct,  in comparison to $8\%$ of the 190 synonymous mutations,  pooled from 18 lineages that span over multiple time points, with an average CDR3 length of 45 bp. We should emphasize that the mutation frequencies $x$ used for statistics of a gene-family  are  evaluated within their respective lineages but the likelihood ratios  $g(x), \, h(x)$ and their uncertainty estimates are aggregate measures in the given gene family (SI).

To see how these observations generalize at the repertoire level, we quantify the region-specific fraction of beneficial and deleterious  mutations within BCR lineages of distinct gene classes and also the fraction of selected mutations that are impeded by clonal interference  (Fig.~\ref{fig:Fig3}c and  Table~\ref{table:Table1}).  Overall we observe that a substantial fraction of lineages  (aggregated into VJ-gene classes) carry positively selected amino acid mutations in their CDR regions  and negatively selected amino acid mutations in FWRs. We infer that  at least ${\ol \alpha_{{}_{\rm benef.}}} =12\% -30 \% $ of CDR mutations that reach frequency $x=0.8$ are strongly beneficial and ${\ol \alpha_{{}_{\rm del.}}} =16\% -20 \%$ of FWR mutations that reach frequency  $x=0.8$ are strongly deleterious (Table~\ref{table:Table1}); overbars indicate averages over VJ-gene classes. Fig.~\ref{fig:Fig4}  shows the detailed statistics of selected mutations  in each patient and Fig.~S7 shows  the inferred effective selection strengths $\ol s$ for different V-gene classes (Methods, SI). The inferred effective selection strengths within the repertoire  indicate a significantly larger fraction of V-gene classes  to carry positively  selected alleles in  their CDRs as opposed to the over-represented negatively selected alleles in FWRs (Fig.~S7). A similar region-specific selection pattern is evident in healthy individuals (Fig.~S8).\\

 \begin{figure}[t!]
\centering\includegraphics[width=\textwidth]{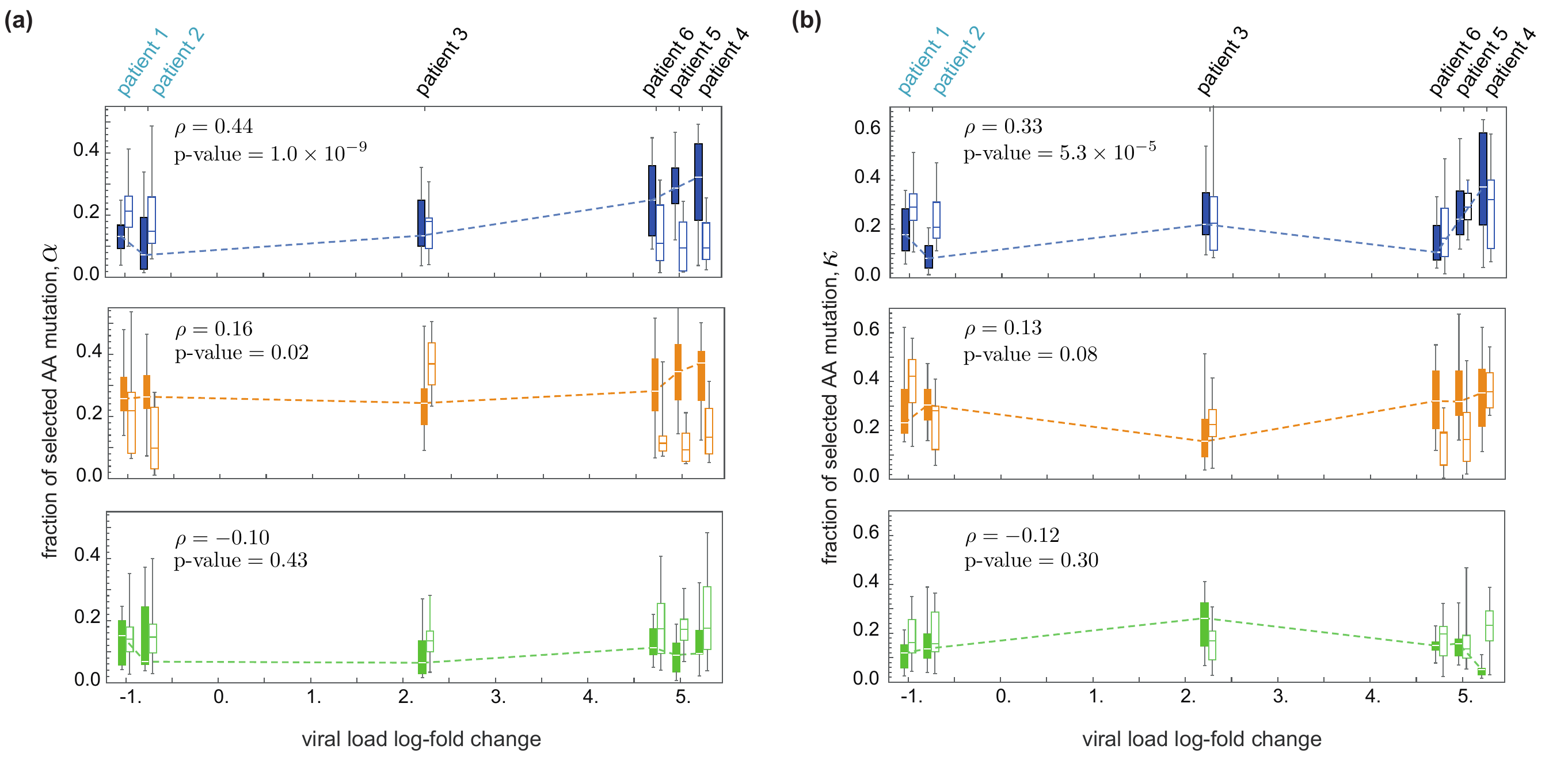}
\caption{\label{fig:Fig4}
{{ {{\bf  Fraction of beneficial and deleterious mutations in BCRs of all patients. }}  {\bf(a)} The box plots show the distributions across V-gene classes for the fraction of beneficial $\alpha_{{}_{\rm benef.}}$ (filled boxes) and deleterious $\alpha_{{}_{\rm del.}}$ (empty boxes) mutations that reach frequency $x=80\%$ within a lineage in each BCR region: CDR3 (top, blue), CDR1/2 (middle, orange) and FWR (bottom, green); mid-line: median; box: 40\% around median; bars: 80\% around median. {\bf(b)} shows similar statistics for the fraction of beneficial and deleterious mutations ($\kappa{{}_{\rm benef.}}$, $\kappa{{}_{\rm del.}}$) that reach frequency $x=60\%$ within a lineage and later go extinct, measuring the extent of clonal interference in B-cells.  The  patients are positioned along the x-axis in accordance to their viral load change $\Delta V$  over the course of infection, estimated as the difference of the averaged log-viral load between the second year ($t>48 \text{ wk}$) and the first year ($t\leq48 \text{ wk}$) of the study, $\Delta  V =  \langle \log V\rangle_{yr. 2} -  \langle \log V\rangle_{yr. 1}$; see Table~S2. Patients 1-2 are ART-naive and the rest had interrupted their treatments after week 48. The dashed line in each figure traces the median  for the distribution of the beneficial mutations (filled boxes) as a function of the change in viral load. In the CDR3 region, positive selection and clonal interference among beneficial mutations are significantly stronger in patients with a larger change in viral load, with Spearman rank correlation and significance indicated in the corresponding panels. The correlations in other regions are small $\sim 0.1$ and  insignificant with ${\rm p-values}>0.01$. Estimated correlations for the fraction of negatively selected mutations are small and insignificant (the values are not shown). Correlations are estimated from the whole data and not only the median of the distributions,  indicated by the dashed lines.
}}}
\end{figure}

\subsection*{Macro- and micro-evolutionary selection fluctuations shape the structure of BCR lineages} 

The frequency-dependent behaviour of the selection likelihood ratio in Fig.~\ref{fig:Fig3} is a strong indicator for the underlying evolutionary mode. In the absence of any competition and clonal interference (i.e., independent site model), the  likelihood  for a beneficial (deleterious) mutation to reach high frequencies should deviate strongly from the neutral expectation, leading to a  rapidly increasing (decreasing) likelihood ratio $g(x)$ as a function of the frequency $x$; see Methods and SI for theoretical expectation in this regime. As shown in Fig.~S6, the data significantly deviates from the expected behaviour of the selection likelihood ratio $g(x)$ for independent site evolution  under  selection. \comment{Competition among beneficial  mutations reduces the rate of BCR adaptation by slowing down the  successive fixation of beneficial mutations and can ultimately hinder the evolution of high affinity BCRs. 
This clonal interference effectively reduces the efficacy of selection~\citep{Schiffels:2011fua} and can lead to flattening of the selection likelihood ratio at high frequencies, consistent with Fig.~\ref{fig:Fig3}.}   Until a secondary competing allele takes up a substantial  fraction of a lineage, the dynamics of the focal allele may be characterized by selection with uncorrelated fluctuations, e.g., \comment{due to spontaneous environmental noise such as access to T-cell help or signalling molecules.} We describe a theoretical model for this process in the Methods and show that the flattening of the likelihood ratio, up to intermediate frequencies, can be explained by an effective selection strength $\ol s$ subject to rapid {\em micro-evolutionary} fluctuations with an amplitude $v$; see Methods, SI and the fitted model to the likelihood ratios in Fig.~S6.  

\comment{The interference likelihood ratio $h(x)$  captures the long-term turnover of circulating alleles, which we characterize by using  the extinction probability of a rising allele  under various  evolutionary scenarios (SI). New beneficial mutations overcome the risk of stochastic extinction by genetic drift once they reach the establishment frequency, which is inversely proportional to their selection coefficient~\citep{Desai:2007wv}. Therefore, for the majority of strongly beneficial mutations at high frequencies, extinction by genetic drift  is an unlikely scenario.  As we discussed above, rapid micro-evolutionary fitness fluctuations due to the environmental noise can slow down the rise of beneficial mutations in a lineage. An  allele which is on  average deleterious  can have a positive selection coefficient at some instances due to fitness fluctuations and intermittently hamper the rise of the dominant beneficial allele. However,  these fluctuations do not persist long enough for a deleterious  mutant  to fully replace an established beneficial allele.  Therefore, neither micro-evolutionary fluctuations nor drift can explain the mutation turnovers  observed in (Figs.~\ref{fig:Fig3},~S6).

We explain the  rise and fall of  beneficial mutations with selection strength $s_0$ by clonal interference where a new beneficial mutation  with selection strength $s_1$ arises  on a distinct  (formerly deleterious or neutral) genetic background and  outcompetes the circulating allele. We model this process as evolution with {\em macro-evolutionary} selection  fluctuations that occur at rates lower than  the life-time of a polymorphism in a lineage~\citep{Mustonen:2009vu}. In this picture, a rising new beneficial allele with selection strength $s_1$ makes a shift in  selection coefficient of the dominant allele $s_0 \to s_0-s_1$. The persistence of such fitness shifts over macro-evolutionary time scales can lead to successive turnover of new beneficial mutations in a lineage, consistent with the observation in Fig.~S6. The comparison between the interference likelihood ratios $h(x)$ predicted  by different evolutionary scenarios strongly indicates the prevalence of clonal interference and macro-evolutionary selection fluctuations in shaping the structure of  BCR lineages for the V-gene classes  in Fig.~S6.}
  
Overall, we observe that the  positively selected  mutations in CDR3 and the pooled CDR1/CDR2 regions are strongly impacted by clonal interference, in contrast to mutations in FWR (Fig.~\ref{fig:Fig3}c, Table~\ref{table:Table1}). {In particular, using the interference likelihood ratio $h(x)$, we infer a significant macro-evolutionary turnover in  the  preference for the selected alleles (see Methods, SI).  Fig.~S9 shows that  strongly selected  alleles  in a lineage are  often replaced  by a competing allele, with a selection strength larger than expected.} These observations indicate the {abundance of beneficial mutations, leading to} pervasive clonal interference {and a long-term selection turnover} in the regions of the BCR with the most important functional role, at the repertoire level. Importantly, our simulations of affinity maturation show that  the inference of  region-specific selection and clonal interference in BCRs is  insensitive to the heterogenous hypermutation statistics and the presence of mutational hotspots in CDRs (Methods and Fig.~S10).

In short, we observe a large fraction of adaptive mutations, and also a substantial amount of clonal interference among them which prevents some of the mutations from dominating within lineages. \\

 \subsection*{Viral expansion drives the BCR repertoire response with strong selection and clonal interference} In patients with interrupted ART, we infer a    
  \comment{substantially} larger fraction of beneficial mutations to rise with strong clonal interference  in pathogen-engaging CDR3 regions following the interruption of treatment, compared to the ART-naive patients with a stable chronic infection (Table~\ref{table:Table1}, Fig.~\ref{fig:Fig3}); The CDR3 statistics of the two patient groups  are significantly distinct based on the two-sample KS-test  for the selection statistics,  $\text{p-value}= 5\times 10^{-8}$ and for the clonal interference  statistics, $\text{ p-value}=5\times 10^{-4}$. Such a shift is not present for mutations in CDR1, CDR2 and FWR  (p-values $>0.1$, two-sample KS-test); see Fig.~\ref{fig:Fig3}, Table~\ref{table:Table1},  SI.  \comment{Moreover, we observe that the  expansion of the HIV population is met with strong positive selection and clonal interference of beneficial mutations in BCRs. Specifically, selection and clonal interference in  the  CDR3 region strongly correlate with changes in viral load during the 2.5 years of study (Fig.~\ref{fig:Fig4},  Fig.~S7 and Table~S2).} No such correlation is observed in  CDR1, CDR2 and FWR (Fig.~\ref{fig:Fig4}). This result is consistent with our inference of strong positive selection and clonal interference in CDR3 of patients who had terminated ART after the first year of treatment, and hence, have the largest change in their viral load.

\begin{table}[h!]
\begin{footnotesize}
\begin{center}
\begin{tabular}{| c || c | c | c | c || c | c | c | c || c | c || c | c |} 
\hline& \multicolumn{4}{|c||}{HIV infected } & \multicolumn{4}{|c||}{HIV infected } & \multicolumn{2}{|c||}{Healthy  } & \multicolumn{2}{|c|}{ Healthy } \\  & \multicolumn{4}{|c||}{untreated} &  \multicolumn{4}{|c||}{interrupted treatment}  &  \multicolumn{2}{|c||}{ productive} & \multicolumn{2}{|c|}{unproductive} \\ \hline 
& ${\ol \alpha_{{}_{\rm benef.}}} $&${\ol\alpha_{{}_{\rm del.}}}$& ${\ol \kappa_{{}_{\rm benef.}}} $&${\ol\kappa_{{}_{\rm del.}}}$&${\ol\alpha_{{}_{\rm benef.}}} $&${\ol\alpha_{{}_{\rm del.}}}$& ${\ol \kappa_{{}_{\rm benef.}}} $&${\ol\kappa_{{}_{\rm del.}}}$& ${\ol\alpha_{{}_{\rm benef.}}}$&${\ol\alpha_{{}_{\rm del.}}}$&${\ol\alpha_{{}_{\rm benef.}}}$&${\ol\alpha_{{}_{\rm del.}}}$\\
 &&&&&&&&&&&&\\ \hline
\multirow{2}{*}{CDR3} & $0.12$ &$0.14$&0.11 & 0.18 & 0.20 & 0.08 & 0.17& 0.12& 0.31 &0.03 & 0.09 & 0.21\\ 
				&$\pm0.01$&$\pm0.01$& $\pm0.01$&$\pm0.01$&$\pm0.01$&$\pm0.01$&$\pm0.01$&$\pm0.01$&$\pm0.01$&$\pm0.004$&$\pm0.01$&$\pm0.01$\\
 \hline
\multirow{2}{*}{CDR1/2}&$ 0.24$& $0.07$& 0.22& 0.11& 0.27 & 0.07 & 0.23& 0.10& \multirow{2}{*}{NA }& \multirow{2}{*}{NA} & \multirow{2}{*}{NA} & \multirow{2}{*}{NA} \\ 
				&$\pm0.01$&$\pm0.01$& $\pm0.02$&$\pm0.01$&$\pm0.01$&$\pm0.01$&$\pm0.01$&$\pm0.01$&&&&\\
\hline
\multirow{2}{*}{FWR} & $0.08$ & $0.13$ &0.09 &0.17 & 0.05 & 0.18 & 0.08& 0.17& 0.07  & 0.22  & 0.11 & 0.24 \\ 
				&$\pm0.01$&$\pm0.01$& $\pm0.01$&$\pm0.01$&$\pm0.01$&$\pm0.01$&$\pm0.01$&$\pm0.01$&$\pm0.01$&$\pm0.01$&$\pm0.01$&$\pm0.01$\\
 \hline
\end{tabular}
\end{center}
\end{footnotesize}
\caption{ {{\bf  Beneficial and deleterious mutations in BCRs.} The average fraction of beneficial $\ol\alpha_{{}_{\rm benef.}}$  and deleterious $\ol\alpha_{{}_{\rm del.}}$ mutations that reach frequency $x=80\%$ (based on selection likelihood ratio $g(0.8)$)  in different regions of BCRs are reported for HIV patients (with interrupted and without treatment) and for healthy individuals (productive and unproductive lineages); averages are estimated over VJ-gene classes.  Similarly, the average fraction of beneficial $\ol\kappa_{{}_{\rm benef.}}$  and deleterious $\ol\kappa_{{}_{\rm del.}}$ mutations that reach frequency $x=60\%$ followed by extinction (based on interference likelihood ratio $h(0.6)$) are reported for HIV patients with interrupted and without treatment; we cannot estimate the interference likelihood ratio  in healthy individuals due to the lack of  time-resolved data. The errors indicate the standard error of the mean. The corresponding distributions are presented in Fig.~\ref{fig:Fig3} and Fig.~S8; see Methods, SI  and  Fig~S5, Fig.~\ref{fig:Fig4} and Fig~S7 for detailed  comparison of selection and clonal interference statistics among the patient groups. {The sequence reads in healthy individuals~\citep{DeWitt:2016cp} are shorter than in HIV patients~\citep{Hoehn:2015kl} and do not extend to  CDR1/2.}  }
\label{table:Table1}
}
\end{table}

This evolutionary  pattern  is consistent with the rate of HIV-1 evolution in patients with different states of therapy. {Genome-wide analysis of HIV-1 has revealed that evolution of the virus within ART-naive patients slows down during  chronic infections  with the majority of mutations happening as reversions rather than immune escape and with limited clonal interference in viral populations~\citep{Zanini:2015gg}. In a separate study  by  SPARTAC~\citep{Roberts:2015},  ART-naive patients show a  slow and steady viral escape from the CTL immune response  over the first 2 years of infection.} Our analysis suggests that the  response at the repertoire level traces the slow evolution of the virus  during the chronic phase. On the other hand, rapid expansion of HIV-1 following the interruption of ART drives a strong immune response. \comment{We hypothesize that evolution of the HIV-1 population during  expansion}  introduces a time-dependent target for the adaptive immune system and opens up room for many beneficial  mutations in the pathogen-engaging CDR3 regions.  The emergence of beneficial mutations on separate backgrounds result in evolution with clonal interference between clones in the repertoire (Figs.~\ref{fig:Fig3},~\ref{fig:Fig4}).   

\section*{Discussion}
Somatic evolution during affinity maturation is complex: there is no one winner of the race for the best antibody.  \comment{We show that the B-cell repertoire mounts a relatively slow response to the stable chronic HIV  during the early stages of infection in ART-naive patients.
 On the other hand, following the interruption of ART in a number of patients, the expanding HIV population drives strong affinity maturation in B-cells with rapid dynamics and clonal interference. Overall, the change in viral load correlates with the  strength of selection and clonal interference in BCR repertoires (Fig.~\ref{fig:Fig4}). Expansion and growth of the HIV-1 population  is often accompanied with rapid evolution of the virus, which can exert a strong selection on the adaptive immune system. We hypothesize that the observed strong selection on BCRs is to counter the viral evolution during its expansion.
The extent of such coevolution can be tested in future experiments that trace the intra-patient dynamics of  BCR and HIV-1 populations over time.}  

The lack of sequence fixation in a repertoire has been previously observed at the level of monoclonal antibody lineages in response to vaccination in mouse models~\citep{Berek:1987tf} and among many rising BCR clones over short time scales  ($\sim$ weeks), during a transient response of human immune repertoires to the influenza vaccine~\citep{Horns:2017kl}. Many factors, including idiotypical interactions, resulting in frequency-dependent selection, or spatial structure of a population have been hypothesized to contribute to the large scale shifts in the repertoire structure, leading to a constant rise of beneficial clones within a repertoire.  Here, we provide a principled approach to characterize  {\em macro-evolutionary} shifts of immune repertoires (Figs.~\ref{fig:Fig3},~\ref{fig:Fig4} and Fig~S9) and show that the somatic evolution of BCRs  is not limited by beneficial mutations,  the supply of which  can last over many years of a chronic infection.

The dynamics of an adaptive immune response  resembles rapid evolution in asexual populations where many beneficial mutations rise to intermediate frequencies  leading to complex {clonal interference} and genetic hitchhiking. Such evolutionary dynamics is prominent in microbial populations~\citep{Lassig:2017hr},   in viruses including HIV  within a patient~\citep{Pandit:2014bh,Zanini:2015gg} and global influenza~\citep{Strelkowa:2012jo,Luksza:2014hj,Neher:2014eu}. {In this evolutionary regime different beneficial mutations arise at nearly the same time and compete with each other, reducing the rate at which beneficial mutations can accumulate. This is distinct from selection, which is merely a difference in growth rate or survival of different cells. Clonal interference can result from competition of BCRs for the same antigen or other stimulatory or activation factors.  In HIV patients, we expect that as long as the CD4+ T-cell levels stay at its normal  range (500 -1500 per ml) to activate a large population of  B-cells,  as it is the case in this study (Table~S2), clonal interference among many positively selected mutations which chase the viral evolution should remain the prominent mode of somatic  evolution in  BCRs on long time-scales.}

On one hand, the nonlinear clonal competition among BCRs reduces \comment{ the rate of adaptation during affinity maturation,} leading to a less predictable fate for good clones, as they can be outcompeted by new mutants before they dominate the immune response. On the other hand, in this evolutionary regime, the fate of a lineage is not strongly impacted by  stochastic uncertainties due to waiting for  new beneficial mutations to arise, as a large supply of such  mutations is available in response to a pathogen. Thus, we hypothesize that it should be feasible to infer {\em fitness models} that primarily rely on the selection differences among  circulating BCRs to forecast the outcome of an immune response. Similar  approaches have been previously successful in forecasting the  fate of a selection-dominated evolving process  to predict the annually dominant strain of the influenza virus~\citep{Neher:2014eu,Luksza:2014hj} and the response of evolving tumors to cancer immunotherapy~\citep{Luksza:2017fg}. Predicting the outcome and efficacy of B-cell response  is of  significant consequence for designing targeted immune-based therapies. Currently, the central challenge in HIV vaccine research is to devise a means to stimulate a lineage producing highly potent broadly neutralizing antibodies (BnAbs). A combination of successive immunization and ART has been suggested as an approach to elicit a stable and effective BnAb response; see e.g. ref.~\citep{Caskey:2016kk}. An optimal treatment strategy should leverage the information on the selected clones among  BCRs during a rapid immune response to antigen stimulation, to overcome the nonlinear impact of clonal interference and derive the immune response towards a desired BnAb within the repertoire.

\section*{Methods}
\noindent{\bf B-cell repertoire data, annotation and genealogies.} We analyze B-cell repertoire data from 6 HIV patients from ref.~\citep{Hoehn:2015kl} with raw sequence reads accessible from the European Nucleotide Archive under study accession numbers, ERP009671 and ERP000572. The data covers $\sim 2.5$ years of study with 6-8 sampled time points per patient; see  Table~S1 for details. The B-cell repertoire sequences consist of 150bp non-overlapping paired-end reads (Illumina MiSeq), with one read covering much of the V gene and the other read covering the area around the CDR3 region and the J gene.  We analyze memory B-cell repertoire data of 3 individuals published in ref.~\citep{DeWitt:2016cp}.  We annotate the  BCR repertoire sequences of each individual (pooled time points) by Partis (version 0.11.0)~\citep{Ralph:2016dr} and further process by MiXCR~\citep{Bolotin:2015jb,Bolotin:2017co}. To identify BCR lineages, we first group sequences by the assigned V gene, J gene and CDR3 length, and then used single linkage clustering  with a threshold of  90\% Hamming distance. We reconstruct a maximum-likelihood  genealogical tree for sequences in each lineage. We use FastTree~\citep{Price:2010eg} to construct the initial tree by maximum parsimony. We use this tree as seed for the maximum likelihood construction of the phylogeny with RAxML~\citep{Stamatakis:2014dj}, using the GTRCAT substitution model.  Details of data processing and error corrections and genealogy reconstruction are discussed in the SI and Fig.~S1.  

\noindent{\bf Selection and interference likelihood ratio.}
Hypermutations during affinity maturation create new clades within a lineage. The frequency $x$ of these clades change over time (Fig.~\ref{fig:Fig3}a). A mutation under positive (or negative) selection should reach a higher (lower) frequency than a neutral mutation. We quantify the likelihood of a mutation reaching a  frequency $x$  in its lifetime by $G(x)\equiv n(x)/N$,  where $n(X)$ is the number of mutations that reach frequency $x$ and $N$ is the  total number of  mutations. We determine the  selection likelihood ratio between non-synonymous $G(x)$ an synonymous $G_0(x)$ mutations,
 \[g(x)=\frac{G(x)}{G_0(x)}\]
We characterize clonal interference  by the likelihood that a mutation reaches frequency $x$ and later goes extinct,  $H(x) = G(x) \times G(0|x)$~\citep{Strelkowa:2012jo}; $G(0|x)$ is the conditional probability that  a mutation starting at  frequency $x$ goes extinct (Fig.~\ref{fig:Fig3}a).  We estimate the interference likelihood ratio by comparing  the interference likelihood between   non-synonymous $H(x)$ and  synonymous mutations, $H_0(x)$  (Fig.~\ref{fig:Fig3}a),
\[ h(x)=\frac{H(x)}{H_0(x)} \equiv \frac{G(x) \times G(0|x)}{G_0 (x) \times G_0(0 | x)}.\]

\noindent{\bf Affinity maturation with  fluctuating selection.}
For independently occurring mutations, the conditional probability  $G(x|x_i,t)$ that a mutation with a starting frequency $x_i $ reaches a frequency $x $ by time $t$ satisfies the backward Kimura's equation~\citep{Kimura:1964},
\[
\frac{\partial}{\partial t} G(x |x_i,t ) = \frac{1}{2N} \left[x(1-x)\frac{\partial^2}{\partial x^2}+ s(t) x(1-x) \frac{\partial}{\partial x} \right] G(x|x_i,t),
\]
with the boundary conditions $G(x|0)=0$ and $G(x_i|x_i)=1$. Here, $s(t)=\overline s+\eta(t)$ is a fluctuating selection coefficient with  average $\ol s$ and uncorrelated Gaussian fluctuations  $\eta(t)$,   with amplitude $v$, which we interpret   as {\em micro-evolutionary}  fluctuations, as opposed to {\em macro-evolutionary} (i.e., long term) correlations in environmental fluctuations. The solution to the stationary state  follows~\citep{Takahata:1975ug}, 
\[
G(x|x_i;\ol s,v) = \frac{1- \left | \frac{1- x_i/a_+}{1-x_i/a_- }\right|^{\lambda(\ol s,v)}}{1- \left | \frac{1- x/a_+}{1- x/a_- }\right|^{\lambda(\ol s,v)}},\]
with $a_{\pm}=\frac{1}{2}[1\pm \sqrt{1+4/v}]$  and $\lambda(\ol s ,v)= \ol s / ( v \sqrt{1+ 4/v})$. Expected selection likelihood ratio with micro-evolutionary fluctuations $g(x;\ol s,v) =  G(x| \ol{x_i} ;\ol s,v) \Big / G(x| \ol{x_i} ;0,v) $ fits the data up to intermediate frequencies $x<0.5$ (Fig. S6), indicating the prevalence of long-term fluctuations in the system. We assume a simple scenario with macro-evolutionary shift in selection preference, where the competing allele can become more beneficial over time, resulting in interference likelihood, 
\[H(x|\overline{x_i};\ol{s_0},{ \ol{s_1}},v)=G(x|\overline x_i;\ol{s_0},v) \, G(1| 1-x;\ol{s_1},v)\]
where $s_1 \neq - s_0$. The expected likelihood ratio $h(x;\ol {s_0},\ol {s_1},v) =  H(x| \ol{x_i} ;\ol {s_0},\ol {s_1},v) \Big / G(x| \ol{x_i} ;0,v) $ fits the data (Fig.~S6). Fig.~S9 shows the prevalence of such macro-evolutionary fluctuations throughout the repertoire.

\section*{Acknowledgments}
We are thankful to Duncan Ralph and Eric Matsen for help with implementation of the BCR analysis software Partis, and Oliver Pybus and Kenneth Hoehn for assisting us with access to the BCR data from HIV patients.   AN acknowledges the funding from the Lewis-Sigler Institute for Integrative Genomics at Princeton University, where this work was initiated. We acknowledge the support by Max Planck Society (AN), SFB1310 (AN and AMW), National Institute of Health grant T32AI055400 (JO), the Janssen Research \& Development LLC (M\L),   ERCCoG grant no. 724208 (TM and AMW). This work was performed in part at the Aspen Center for Theoretical Physics, which is supported by the National Science Foundation grant PHY-1066293.

\section*{Competing interests}
The authors declare no competing interests.

\clearpage{}
\newpage{}

\clearpage{]
\newpage{}



\section*{Supplementary material}

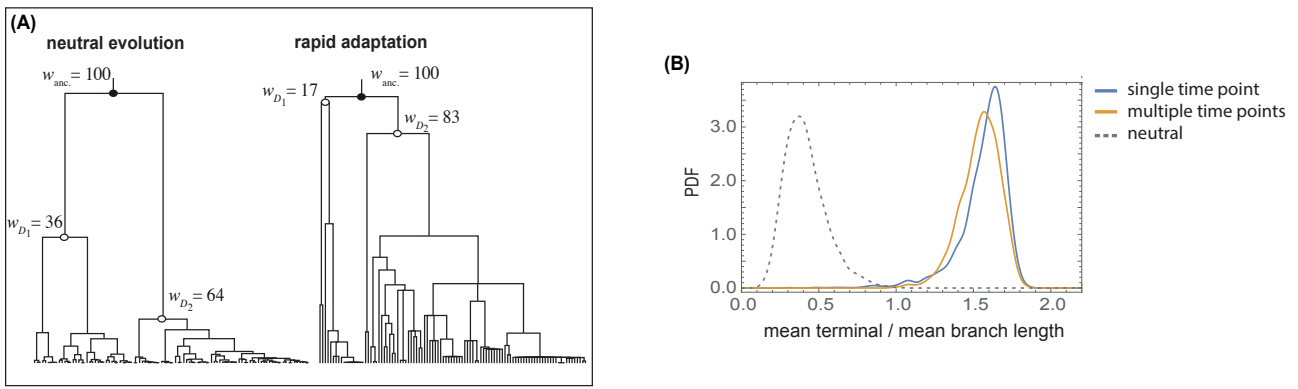

Figure S1: **Impact of selection on lineage tree statistics.** **(A)** Simulated phylogenies for neutral evolution (left) and rapid adaptation with positive selection (right), generated by the coalescence package [29] and plotted with FastTree [9]. The weights of the ancestral node $w_{\text{anc.}}$ and its daughters $w_{D_1}$, $w_{D_2}$ (i.e., their clone size) are indicated in each phylogeny. **(B)** Branch length statistics for lineages sampled in only one time point from HIV patients. The distribution of mean terminal branch length (in units of divergence time) relative to the mean length of all branches is comparable between BCR lineages of HIV patients that are present in only a single time point (blue) and lineages sampled over multiple time points (orange). The corresponding distribution for the simulated neutral trees (similar to Fig. 2B) is shown by a dotted line. The elongated terminal branches in BCR lineages is indicative of positive selection.

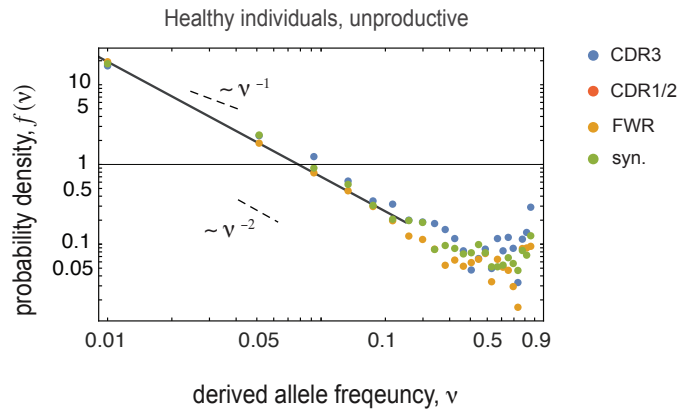

Figure S2: **Site frequency spectrum of the unproductive BCR lineages.** SFS $f(\nu)$ is shown for mutations in different regions of BCRs (distinct colors) in unproductive lineages of healthy individuals; see Fig. 2C for SFS of productive lineages.

7

**HIV, untreated**

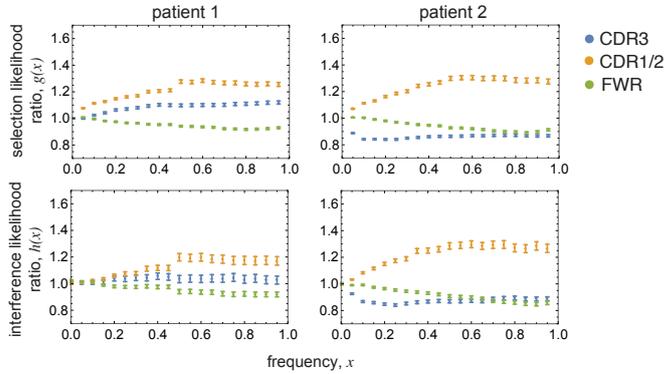

**HIV, interrupted ART**

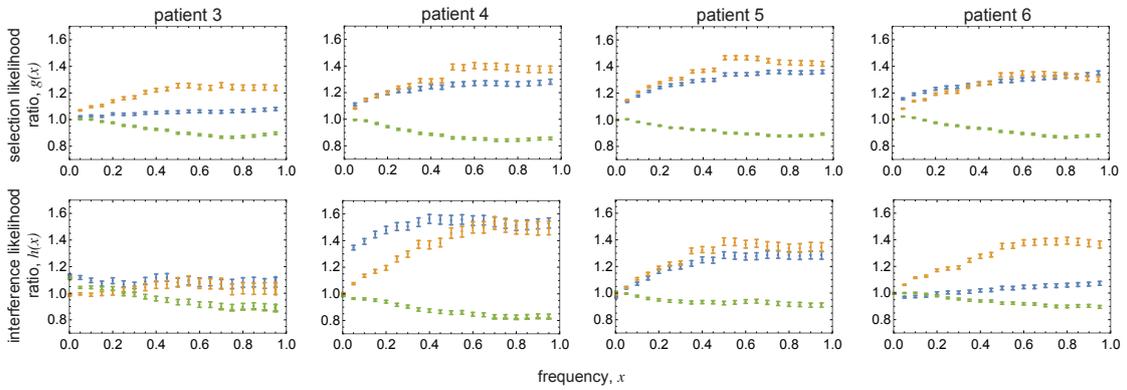

**Healthy, productive**

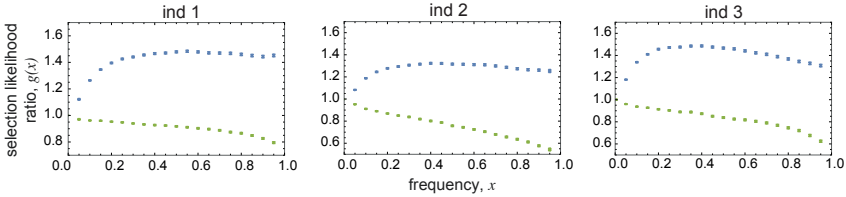

**Healthy, unproductive**

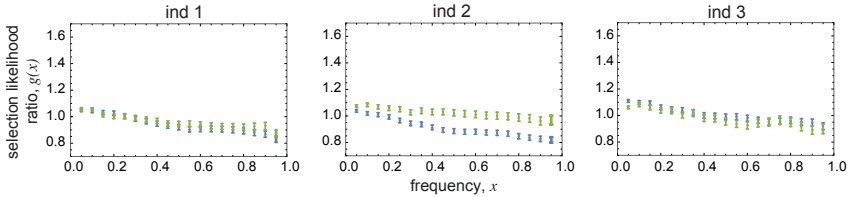

Figure S3: **Selection and interference likelihood ratios in all individulas.** Panels show selection and interference likelihood ratios $g(x)$, $h(x)$ in HIV patients (untreated and with interrupted ART) and the selection likelihood ratio $g(x)$ in productive and unproductive lineages of healthy individuals, estimated from all lineages in each individual. We consistently see strong evidence for negative selection in FWR regions of productive lineages and positive selection in both or either of CDR regions. We do not see such distinction in unproductive lineages. Note that the repertoire level averaged likelihood ratios are highly coarse grained statistics and miss the gene-specific evidence for selection and clonal interference, as shown in Fig. 3.

8

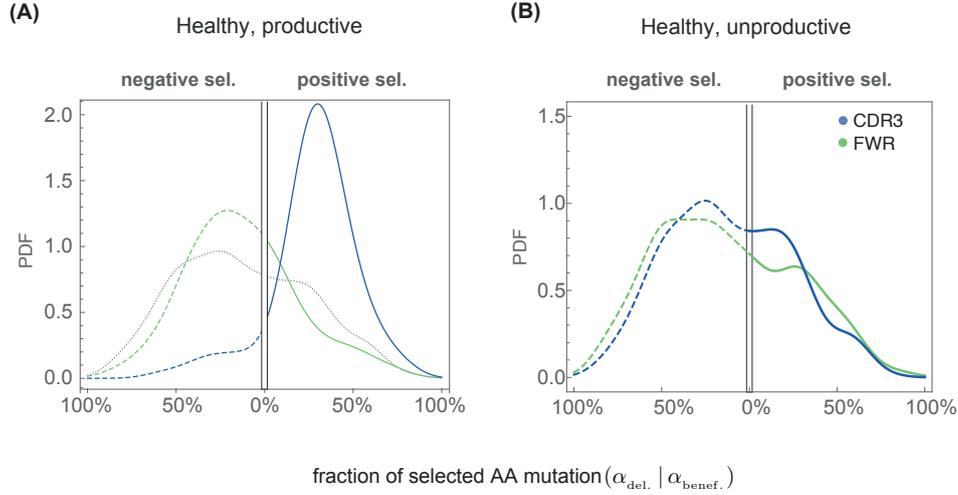

Figure S4: **Fraction of selected BCR mutations in healthy individuals.** The probability density across distinct VJ-gene classes for the (minimum) fraction of beneficial (right) $\alpha_{\text{benef.}}$ and deleterious (left; inverted x-axis) $\alpha_{\text{del.}}$ amino acid changes that reach frequency $x = 80\%$ within a lineage is shown for different regions of BCRs in **(A)** productive and **(B)** unproductive lineages of healthy individuals. Similar to HIV patients (Fig. 3), the CDR3 mutations in productive lineages of healthy individuals are under positive selection, whereas the FWR mutations are under negative selection. We do not infer any significant differences between selection patterns in CDR3 and in FWR region of unproductive lineages. The probability density for mutations pooled from both regions of unproductive lineages is shown as the null expectation (dotted gray line), similar to Fig. 3.

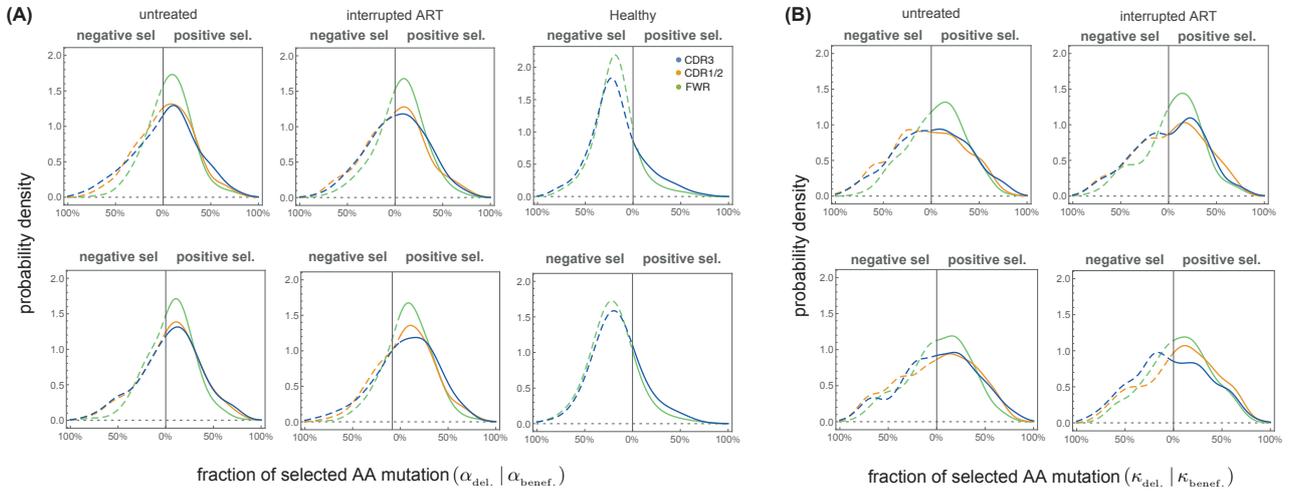

Figure S5: **Robustness of selection inference to BCR hypermutation biases.** The figure shows the statistics of selected mutations for simulated neutral hypermutation processes along the inferred B-cell lineages, as described in Section 5 of SI. We use two hypermutation models, IGoR statistics [28] (top row) and the S5F model [24] (bottom row). Similar to Fig. 3, each panel shows the probability density across distinct VJ-gene classes for **(A)** the minimum fraction of beneficial / deleterious mutations ($\alpha_{\text{benef.}} / \alpha_{\text{del.}}$) that reach frequency $x = 80\%$, and **(B)** for beneficial / deleterious mutation fractions ($\kappa_{\text{benef.}} / \kappa_{\text{del.}}$) that reach frequency $x = 60\%$ and later go extinct. These statistics are estimated for mutations in different regions of BCRs (colors) in healthy individuals, HIV patients with interrupted ART and in untreated HIV patients. The region-specific pattern of selection seen in Fig. 3 is simulated lineages with context-dependent hypermutation models.

9

|  |  | HIV infected (untreated) | | HIV infected (interrupted ART at week 48) | | | | Healthy | | |
|---|---|---|---|---|---|---|---|---|---|---|
|  |  | patient 1 | patient 2 | patient 3 | patient 4 | patient 5 | patient 6 | ind 1 | ind 2 | ind 3 |
| # productive lineages | size | | | | | | | | | |
|  | $> 20$ | 3,335 | 3,702 | 3,164 | 2,125 | 4,342 | 4,155 | 21,486 | 15,755 | 15,782 |
|  | $> 50$ | 785 | 773 | 613 | 485 | 1,246 | 1,205 | 5,242 | 3,978 | 3,390 |
| # unproductive lineages | $> 20$ | 0 | 0 | 0 | 0 | 0 | 0 | 897 | 1041 | 962 |
|  | $> 50$ | 0 | 0 | 0 | 0 | 0 | 0 | 177 | 198 | 155 |
| time samples (weeks) |  | 0, 4, 16, 24, 52, 72, 120 | 0, 4, 12, 16, 24, 52 60, 108 | 4, 12, 16, 24, 52, 60, 108 | 4, 12, 16, 24, 52, 60, 108 | 4, 12, 16, 24, 52, 60, 108 | 0, 4, 16, 24, 52, 60, 108 | 0 | 0 | 0 |

Table S1: Statistics of reconstructed BCR lineages with size ($> 20$) and ($> 50$) and the sampled time points after the start of the study in HIV infected patients and in healthy individuals.


\end{document}